\documentclass[12pt]{article}

  \usepackage{amsfonts}
\begin{document}

\begin{center}
{\large \bf Topological origin  of vector boson masses in  Electroweak Model}
\end{center}

\begin{center}
N.A.~Gromov \\
Department of Mathematics, Komi Science Center UrD, RAS, \\
Kommunisticheskaya st. 24, Syktyvkar 167982, Russia \\
E-mail: gromov@dm.komisc.ru
\end{center}

\begin{abstract}
Starting from   the fact that the quadratic form $\phi^{\dagger}\phi=R^2$ of the matter field space $C_2$ of the Electroweak Model is invariant with respect to gauge transformations we have suggested to regard fields on the compact 3-dimensional sphere $S_3$, which is defined by this quadratic form,  instead of noncompact space $C_2$ or $R_4$ if real components are counted.
The vector boson masses are automatically generated by   the {\it free}  Lagrangian  on the sphere $S_3$ and are  expressed through the sphere radius $R$.
 Higgs boson field does not presented in the model.

\end{abstract}

\vspace{3mm}

\section{Introduction}

The Standard Electroweak Model  based on gauge group $ SU(2)\times U(1)$ gives a good  description of electroweak processes.   The massive vector bosons predicted by the model was experimentally observed and have the masses $m_W=80GeV $ for charged W-boson and $m_Z=91GeV$ for neutral Z-boson.
  One of   the unsolved problems is the origin of the vector bosons masses.
In the standard formulation these masses are arisen   as a result of
spontaneous symmetry breaking vie Higgs mechanism which
include three steps:
1) the potential of the  self-acting scalar field $\phi$ of the special form $V(\phi)=\mu^2\bar{\phi}\phi + \lambda (\bar{\phi}\phi)^2 $ is introduced
{\it by hand} in the Lagrangian;
2) its minimal values are considered for  imaginary mass $\mu^2 <0$ and are interpreted as degenerate vacuum;
3) one of the gauge equivalent vacuum is fixed and then all fields are regarded in the  neighborhood of this vacuum.

Sufficiently  artificial Higgs mechanism  with its imaginary bare mass is a naive relativistic analog of the  phenomenological description of superconductivity \cite{Sh-09}.
Therefore there are a serious doubt whether electroweak symmetry is broken by such a Higgs mechanism, or by something else.
 The existence of the Higgs boson is not yet
experimentally verified.
One expect that the future experiments on LHC will given definite answer on the question: does Higgs boson really exist or not.
 The emergence  of  large number  Higgsless models  \cite{C-05}--\cite{MT-08} was stimulated by difficulties with Higgs boson. These models are mainly based on extra dimensions of different types or  larger gauge groups.
A finite electroweak model without a Higgs particle which is used  a regularized quantum field theory \cite{E-66},\cite{E-67} 
was developed in \cite{MT-08}.

The simple mechanism for generation of the vector boson masses in Electroweak Model was recently suggested \cite{Gr-07}.
It is based on the fact that  the quadratic form $\phi^{\dagger}\phi=R^2$ in the matter fields space $\Phi_2$ is invariant with respect to gauge transformations.
This form  define the 3-dimensional sphere $ S_3$ of the radius $R$ in the space $\Phi_2$ which is  $ C_2$ or $ R_4$ if real components are counted.
The vector boson masses are automatically generated by   the {\it free} (without any potential term) Lagrangian  on the sphere $S_3$ and are  expressed through the sphere radius $R$, so there is no need in special mechanism of spontaneous symmetry breaking. Higgs boson field does not presented in the model.
The free Lagrangian on the sphere $ S_3$ can be obtained as well from standard Electroweak Lagrangian by transformation of the Cartesian coordinates in $\Phi_2$ to a some  coordinates on the sphere $ S_3$.
This transformation corresponds to  the transition from linear to nonlinear representation of the gauge group in the space of functions on $ S_3$.
The  fermion Lagrangian of the Standard Electroweak Model are modified  by replacing of the  fields $\phi$ with the restricted on the quadratic form fields  in such a way that its second order terms provide the electron mass and neutrino remain massless.

\section{Modified  Electroweak Model}

 The  bosonic   part of the Electroweak Model is given by the sum \cite{R-99}
\begin{equation}
L_B=L_A + L_{\phi}.
\label{e1}
\end{equation}
 where
\begin{equation}
L_A=\frac{1}{8g^2}\mbox{Tr}(F_{\mu\nu})^2-\frac{1}{4}(B_{\mu\nu})^2=
-\frac{1}{4}[(F_{\mu\nu}^1)^2+(F_{\mu\nu}^2)^2+(F_{\mu\nu}^3)^2]-\frac{1}{4}(B_{\mu\nu})^2
\label{e8}
\end{equation}
is  the gauge fields Lagrangian for $SU(2)\times U(1)$ group.
 The gauge fields are
$$
A_\mu (x)=-ig\sum_{k=1}^{3}T_kA^k_\mu (x)=-i\frac{g}{2}\left(
\begin{array}{cc}
A^3_{\mu} &  A_\mu^1-iA_\mu^2  \\
A_\mu^1+iA_\mu^2 & -A^3_{\mu}
\end{array}\right)  ,
$$
\begin{equation}
\quad B_\mu (x)=-ig'YB_\mu (x)=-i\frac{g'}{2}\left(
\begin{array}{cc}
B_{\mu} &  0  \\
0 & B_{\mu}
\end{array}\right),
\label{dop-1}
\end{equation}
where  $g,g'$ are coupling constants,
 $T_k=\frac{1}{2}\tau_k$, with $\tau_k$ being Pauli matrices,
take their values in Lie algebras $su(2),$  $u(1)$ respectively.
 The stress tensors  look as follows
\begin{equation}
F_{\mu\nu}(x)={\cal F}_{\mu\nu}(x)+[A_\mu(x),A_\nu(x)],\quad
{\cal F}_{\mu\nu}=\partial_{\mu}A_{\mu}-\partial_{\nu}A_{\mu},
\quad
 B_{\mu\nu}=\partial_{\mu}B_{\nu}-\partial_{\nu}B_{\mu}.
\label{dop-2}
\end{equation}

The second term in (\ref{e1})
\begin{equation}
  L_{\phi}= \frac{1}{2}(D_\mu \phi)^{\dagger}D_\mu \phi
\label{e9}
\end{equation}
is the {\it free} matter fields Lagrangian.
Here $D_{\mu}$ are the covariant derivatives
\begin{equation}
D_\mu\phi=\partial_\mu\phi -ig\left(\sum_{k=1}^{3}T_kA^k_\mu \right)\phi-ig'YB_\mu\phi
\label{e5}
\end{equation}
of the fields $\phi$ from
 the space $C_2$  of fundamental representation of $ SU(2)$.

For  $\Omega\in SU(2),\; e^{i\omega}\in U(1)$ the gauge transformations of the fields are as follows
$$
\phi^{\Omega}=\Omega\phi, \quad \phi^{\omega}=e^{i\omega}\phi,
$$
$$
A_\mu^{\Omega}(x)= \Omega A_\mu(x)\Omega^{-1} -\partial_{\mu}\Omega\cdot \Omega^{-1}, \quad A_\mu^{\omega}(x)= A_\mu(x),
$$
\begin{equation}
B_{\mu}^{\Omega}= B_{\mu}, \quad B_{\mu}^{\omega}= B_{\mu}- 2\partial_{\mu}\omega
\label{eq9}
\end{equation} 
and  Lagrangian  (\ref{e1})  is invariant under $SU(2)\times U(1)$ gauge group.

The Lagrangian   $L_B=L_A+L_{\phi}$  describe   massless fields.
In a standard approach to generate   mass terms for  vector bosons   the Higgs mechanism is used.
 We have proposed \cite{Gr-07} new mechanism  for generation masses of  vector bosons
 in Electroweak Model which is based on the topological idea of the restriction of the free bosonic Lagrangian $L_B=L_A + L_{\phi} $ from the whole noncompact space $ R_4$ to the compact sphere $ S_3$.
The complex two dimensional matter fields space $C_2$ can be regarded as four dimensional real Euclidean space $ R_4$ if real components are counted.
 The quadratic form
\begin{equation}
  \phi^\dagger\phi=\phi_1^*\phi_1 + \phi_2^*\phi_2=R^2
\label{d1}
\end{equation}
  is invariant with respect to $ SU(2)\times U(1)$ group and define the three dimensional  sphere $  S_3 $ in $C_2$. Therefore the restriction of the Electroweak  Model on the sphere $ S_3 $ will be  the gauge model with the same  gauge symmetry. 
Similar restriction (\ref{d1}) was appeared in a unified conformal model for fundamental interactions \cite{PR-94} as a consequence of the gauge fixing freedom connected with the local conformal symmetry group.

   Let us introduce the  real fields
   $ r,\; \bar{\psi}=(\psi_1,\psi_2,\psi_3)$
  by the equations
\begin{equation}
\phi_1=r(\psi_2+i\psi_1), \quad \phi_2=r(1+i\psi_3),
\label{dop-3}
\end{equation}
   then quadratic form (\ref{d1}) is written as
\begin{equation}
r^2(1 + \bar{\psi}^2)=R^2,
\label{dop-4}
\end{equation}
where $\bar{ \psi}^2=\psi_1^2+ \psi_2^2+\psi_3^2$, and define the 
sphere $  S_3 $  of the radius $R >0$ in the  space  $ R_4$.
Three independent real fields $\bar{\psi}$ form  intrinsic Beltrami coordinate system on $  S_3 $
  with the following  metric tensor
\begin{equation}
g_{kk}(\bar{\psi})=\frac{1+\bar{\psi}^2-\psi_k^2}{(1+\bar{\psi}^2)^2}, \;\;
g_{kl}(\bar{\psi})=\frac{-\psi_k \psi_l}{(1+\bar{\psi}^2)^2}, \; k,l=1,2,3.
\label{dop-5}
\end{equation}
Let us define the {\it free}   matter fields Lagrangian $L_\phi$  with the help of the metric tensor  of  the spherical space $ S_3 $ in the form 
\begin{equation}
L_\psi=\frac{R^2}{2}\sum_{k,l=1}^3g_{kl}D_\mu\psi_kD_\mu\psi_l=
\frac{R^2\left[(1+\bar{\psi}^2)(D_{\mu}\bar{\psi})^2-(\bar{\psi},D_{\mu}\bar{\psi})^2\right] }{2(1+\bar{\psi}^2)^2}.
\label{d7}
\end{equation}
The covariant derivatives   are obtained with the help of  the standard expressions  (\ref{e5}), using now the nonlinear representations of generators for the algebras $su(2),$ $u(1)$ in the space $S_3$ \cite{Gr-07}
$$
T_1\bar{\psi}=\frac{i}{2}\left(\begin{array}{c}
-(1+\psi_1^2)	 \\
 \psi_3-\psi_1\psi_2\\
 -(\psi_2+\psi_1\psi_3)
\end{array} \right), \quad
T_2\bar{\psi}=\frac{i}{2}\left(\begin{array}{c}
-(\psi_3+\psi_1\psi_2) \\
 -(1+\psi_2^2)\\
 	\psi_1-\psi_2\psi_3
\end{array} \right),
$$
\begin{equation}
T_3\bar{\psi} =\frac{i}{2}\left(\begin{array}{c}
-\psi_2+\psi_1\psi_3	\\
 \psi_1+\psi_2\psi_3 \\
 1+\psi_3^2
\end{array} \right), \quad
Y\bar{\psi} =\frac{i}{2}\left(\begin{array}{c}
-(\psi_2+\psi_1\psi_3)	\\
 \psi_1-\psi_2\psi_3 \\
 -(1+\psi_3^2)
\end{array} \right)
\label{dop-6}
\end{equation}
and are as follows:
$$
D_\mu \psi_1=\partial_\mu \psi_1+\frac{g}{2}\left[-(1+\psi_1^2)A_\mu^1 -(\psi_3+\psi_1\psi_2)A_\mu^2-(\psi_2-\psi_1\psi_3)A_\mu^3 \right] - \frac{g'}{2}(\psi_2+\psi_1\psi_3)B_\mu,
$$
$$
D_\mu \psi_2=\partial_\mu \psi_2+\frac{g}{2}\left[(\psi_3-\psi_1\psi_2)A_\mu^1 -(1+\psi_2^2)A_\mu^2 + (\psi_1+\psi_2\psi_3)A_\mu^3 \right] + \frac{g'}{2}(\psi_1-\psi_2\psi_3)B_\mu,
$$
\begin{equation}
D_\mu \psi_3=\partial_\mu \psi_3+\frac{g}{2}\left[-(\psi_2+\psi_1\psi_3)A_\mu^1 +(\psi_1-\psi_2\psi_3)A_\mu^2+(1+\psi_3^2)A_\mu^3 \right] - \frac{g'}{2}(1+\psi_3^2)B_\mu.
\label{eqn11}
\end{equation}
Let us stress that Lagrangian (\ref{d7}) can be obtain  from  the Lagrangian (\ref{e9}) by the transformation from the Cartesian coordinates in Euclidean space $R_4$ to the Beltrami coordinates on the sphere $ S_3$, when sphere radius does not depend on space-time variables.
The gauge fields Lagrangian (\ref{e8}) does not depend on the fields $\phi$ and therefore remains unchanged.
So the full Lagrangian $L_B$ is given by the sum of (\ref{e8}) and (\ref{d7}).

The particle content of the model is described by the second order part of the full  Lagrangian.
The ground state of  $L_B$  correspond to zero fields.
For  small fields in the neighborhood of ground state, the second order part of  $L_\psi $  is written as
\begin{equation}
L_\psi^{(2)}=
\frac{R^2}{2}\sum_{k=1}^3\left[(D_\mu \psi_k)^{(1)}\right]^2 ,
\label{eqn-1}
\end{equation}
 where linear terms $(D_\mu \psi_k)^{(1)} $ in covariant derivatives  have the form
$$
(D_\mu \psi_1)^{(1)}=  -\frac{g}{2}\left(A_\mu^1-\frac{2}{g}\partial_\mu\psi_1\right) =-\frac{g}{2}\hat{A}_\mu^1,
$$
$$
(D_\mu \psi_2)^{(1)}=
-\frac{g}{2}\left(A_\mu^2-\frac{2}{g}\partial_\mu\psi_2\right)= -\frac{g}{2}\hat{A}_\mu^2, 
$$
$$ 
(D_\mu \psi_3)^{(1)}=
\partial_\mu\psi_3+\frac{1}{2}(gA_\mu^3-g'B_\mu)=\frac{1}{2}\sqrt{g^2+g'^2}Z_\mu.
$$ 
For the new fields
$$
W^{\pm}_\mu = \frac{1}{\sqrt{2}}\left(\hat{A}^1_\mu \mp i \hat{A}^2_\mu \right), \quad (W^{-}_\mu)^*=W^{+}_\mu
$$
\begin{equation}
 Z_\mu =\frac{gA^3_\mu-g'B_\mu + 2\partial_\mu \psi_3}{\sqrt{g^2+g'^2}}, \quad
 A_\mu =\frac{g'A^3_\mu+gB_\mu}{\sqrt{g^2+g'^2}}
\label{e24}
\end{equation}
the quadratic  part of the full Lagrangian
$$
L_B^{(2)}= L_A^{(2)} + L_{\psi}^{(2)}=
$$
\begin{equation}
= -\frac{1}{2}{\cal W}^{+}_{\mu\nu}{\cal W}^{-}_{\mu\nu} + m_{W}^2W^{+}_\mu W^{-}_\mu - \frac{1}{4}{\cal F}_{\mu\nu}{\cal F}_{\mu\nu}
  - \frac{1}{4}{\cal Z}_{\mu\nu}{\cal Z}_{\mu\nu} +\frac{m_Z^2}{2}Z_\mu Z_\mu,
\label{eqn-2}
\end{equation}
where ${\cal X}_{\mu\nu}=\partial_{\mu}X_{\nu}-\partial_{\nu}X_{\mu} $ for
$ X_{\mu}=W^{\pm}_{\mu},Z_{\mu}$,
 describes massive vector fields  $W_\mu^{\pm}$ with identical mass  $m_W=\frac{1}{2}gR$ ( $W$-bosons),
  massless  vector field $A_\mu $ (photon)
and massive vector field     $Z_\mu,\;$    $m_Z=\frac{R}{2}\sqrt{g^2+g'^2}$ ( $Z$-boson).
In other words it describes all the experimentally verified parts of the  Electroweak Model,   but  does not include the scalar Higgs field.
The particle masses  are identical to those of the Standard  Model
and are expressed  through the free parameter $R$ of the model, which is now interpreted as the curvature radius of the spherical matter field space $S_3$.

The fermion Lagrangian of the standard Electroweek Model
is taken in the form \cite{R-99}
\begin{equation}
L_F=L_l^{\dagger}i\tilde{\tau}_{\mu}D_{\mu}L_l + e_r^{\dagger}i\tau_{\mu}D_{\mu}e_r -
h_e[e_r^{\dagger}(\phi^{\dagger}L_l) +(L_l^{\dagger}\phi)e_r],
\label{eq14}
\end{equation}
where
$
L_l= \left(
\begin{array}{c}
	e_l\\
	\nu_{e,l}
\end{array} \right)
$
is the $SU(2)$-doublet,  $e_r $ the $SU(2)$-singlet, $h_e$ is constant and $e_r, e_l, \nu_e $ are two component Lorentz spinor.
Here $\tau_{\mu}$ are Pauli matrices,
$\tau_{0}=\tilde{\tau_0}={\bf 1},$ $\tilde{\tau_k}=-\tau_k. $
The covariant derivatives $D_{\mu}L_l $ are given by (\ref{e5}) with $L_l$ instead of
$\phi$ and $D_{\mu}e_r=(\partial_\mu + ig'B_\mu)e_r. $
 The convolution on the inner indices of $SU(2)$-doublet is denoted by $(\phi^{\dagger}L_l)$.

The matter field $\phi$  appears in Lagrangian (\ref{eq14}) only in mass terms.
After expression $\phi$ through $\bar{\psi}$
 the mass terms are rewritten in the form
$$
h_e[e_r^{\dagger}(\phi^{\dagger}L_l) +(L_l^{\dagger}\phi)e_r]=
\frac{h_eR}{\sqrt{1+\bar{\psi}^2}}\left\{e_r^{\dagger}e_l + e_l^{\dagger}e_r + \right.
$$
\begin{equation}
\left.  +i\psi_3\left(e_l^{\dagger}e_r - e_r^{\dagger}e_l \right) +
i \left[\psi_1\left(\nu_{e,l}^{\dagger}e_r - e_r^{\dagger}\nu_{e,l} \right)+
i\psi_2\left(\nu_{e,l}^{\dagger}e_r + e_r^{\dagger}\nu_{e,l} \right)\right]
\right\}.
\label{n1-1}
\end{equation}
Its  second order terms
$ h_eR\left(e_r^{\dagger}e_l^{-} + e_l^{- \dagger}e_r \right)$
provide  the electron mass $m_e=h_eR, $ and neutrino  remain massless.

The topological idea with compact matter field space $S_3$ instead of noncompact one $R_4$ was used and developed in several  
papers \cite{F-08}--\cite{ACL-09}.
The transformation of the free Lagrangian from Cartesian to radial coordinates  $ R_{+}\times S_3$ in $R_4$ was regarded in \cite{F-08}, where the sphere $S_3$ was  parametrized by elements of $SU(2)$ groups. 
When sphere radius depend on space-time coordinates
  the real positive massless scalar field is presented  in the model.
  
We are interested in  
 parametrization of $S_3$ with $R= const$ by elements of $SU(2)$ groups, so
write $\phi$ as
\begin{equation}
\phi=\left(\begin{array}{c}
 \phi_1\\
\phi_2
\end{array} \right)=R\left(\begin{array}{c}
 \chi_1\\
\chi_2
\end{array} \right)=R\left(\begin{array}{cc}
 \chi_1 & -\chi^*_2\\
\chi_2 & \chi^*_1
\end{array} \right) 
\left(\begin{array}{c}
 1\\
0
\end{array} \right) \equiv R h \varphi_0,
\label{eq10}
\end{equation} 
then from (\ref{d1})  
it follows that
\begin{equation}
\chi^{\dagger}\chi=\chi^*_1\chi_1 +  \chi^*_2\chi_2=1
\label{eq11}
\end{equation} 
and the matrix $h$ is unimodular $\det h=1$ and unitary $h^{\dagger}h=1.$   
So the vector $\chi= h \varphi_0 \in S_3 $ or the matrix $h\in SU(2)$ defines the coordinates on the sphere (\ref{eq11}).
The matrix $h$ transforms  under the gauge transformations as
\begin{equation}
h^{\Omega}=\Omega h,\quad
h^{\omega}=\left(\begin{array}{cc}
 e^{i\omega}\chi_1 & - e^{-i\omega}\chi^*_2\\
 e^{i\omega}\chi_2 &  e^{-i\omega}\chi^*_1
\end{array} \right)=h e^{i\omega \tau_3}, 
\label{eq12}
\end{equation} 
where $\tau_3= \mbox{diag}(1,-1).$

The matter fields Lagrangian (\ref{e9})   takes the form
\begin{equation}   
  L_{\phi}=   
  \frac{1}{2} \frac{R^2}{4} \left(\sqrt{g^2+g'^2}\right)^2 \left(Z_{\mu}\right)^2 +
  \frac{R^2g^2}{4}W_{\mu}^{+}W_{\mu}^{-,}
\label{eq17}
\end{equation} 
where the new fields are introduced
$$   
W_{\mu}^{\pm}=\frac{1}{\sqrt{2}}\left(W_{\mu}^1 \mp iW_{\mu}^2\right), \quad 
Z_{\mu}=\frac{gW_{\mu}^3 + g'B_{\mu}}{\sqrt{g^2+g'^2}}, \quad 
A_{\mu}=\frac{g'W_{\mu}^3 - gB_{\mu}}{\sqrt{g^2+g'^2}},
$$
\begin{equation} 
W_{\mu}(x)= h^{\dagger}A_{\mu}(x)h + h^{\dagger}\partial_{\mu}h=-i\frac{g}{2}\sum_{k=1}^{3}W_{\mu}^k\tau_k.
\label{eq18}
\end{equation} 
These fields  are invariant under $SU(2)$ transformations: 
$ X^{\Omega}=X,\;\; X=W_{\mu}^{\pm},  Z_{\mu}, A_{\mu} $
and are transforms under those  of $U(1)$ as 
\begin{equation}    
\left(W_{\mu}^{\pm}\right)^{\omega}=e^{\mp2i\omega}W_{\mu}^{\pm}, \quad 
Z_{\mu}^{\omega}=Z_{\mu}, \quad 
A_{\mu}^{\omega}=A_{\mu} + \frac{2}{e}\partial_{\mu}\omega, \quad 
e=\frac{gg'}{\sqrt{g^2+g'^2}}.
\label{eq19}
\end{equation} 
The stress tensors of the fields $A_{\mu}(x)$ and $W_{\mu}(x)$ (\ref{eq18}) are connected by 
\begin{equation}
W_{\mu\nu}(x)=h^{\dagger}F_{\mu\nu}(x)h 
\label{eq20}
\end{equation} 
therefore
$
\mbox{tr}(F_{\mu\nu}(x))^2=\mbox{tr}(W_{\mu\nu}(x))^2 
$
 and the gauge fields Lagrangian (\ref{e8}) can be written as
\begin{equation}
L_A=-\frac{1}{4}\left[(W_{\mu\nu}^1)^2 + (W_{\mu\nu}^2)^2 \right] -\frac{1}{4}\left[(W_{\mu\nu}^3)^2 + (B_{\mu\nu})^2 \right].
\label{eq22}
\end{equation} 
Then
the bosonic Lagrangian (\ref{e1}) assumes the form
$$
L_B=   -\frac{1}{2}{\cal{W}}^{+}_{\mu\nu}{\cal{W}}^{-}_{\mu\nu}  + m_W^2 W_{\mu}^{+}W_{\mu}^{-} 
 -\frac{1}{4} ({\cal F}_{\mu\nu})^2 -\frac{1}{4}({\cal{Z}}_{\mu\nu})^2 
 + \frac{1}{2} m_Z^2  \left(Z_{\mu}\right)^2 +
 $$
\begin{equation}
+ \frac{igP - g^2S - (g{\cal Z}_{\mu\nu} + g'{\cal F}_{\mu\nu})H_{\mu\nu} }{2\sqrt{g^2+g'^2}} 
-\frac{1}{4}(H_{\mu\nu})^2,
\label{eq27}
\end{equation} 
where  terms with third and fourth powers in fields are as follows
$$
P={\cal{W}}^{+}_{\mu\nu}\left[W_{\mu}^{-}(gZ_{\nu}+g'A_{\nu}) - W_{\nu}^{-}(gZ_{\mu}+g'A_{\mu}) \right] - 
$$
$$
 - {\cal{W}}^{-}_{\mu\nu}\left[W_{\mu}^{+}(gZ_{\nu}+g'A_{\nu}) - W_{\nu}^{+}(gZ_{\mu}+g'A_{\mu}) \right],
$$
$$ 
S=\left[W_{\mu}^{+}(gZ_{\nu}+g'A_{\nu}) - W_{\nu}^{+}(gZ_{\mu}+g'A_{\mu}) \right] 
\left[W_{\mu}^{-}(gZ_{\nu}+g'A_{\nu}) - W_{\nu}^{-}(gZ_{\mu}+g'A_{\mu}) \right],
$$
\begin{equation}
H_{\mu\nu}= -ig\left(W_{\mu}^{+}W_{\nu}^{-} - W_{\mu}^{-}W_{\nu}^{+} \right).
\label{e23}
\end{equation} 
The second order terms of (\ref{eq27}) coinside with thouse of (\ref{eqn-2}), i.e. the particle contents of the model are identical.
This 
demonstrate that the generation vector bosons masses do not depend on the  choice of the specific coordinate system on $S_3$, but has topological origin. 
 The parametrization of the sphere $S_3$ by the elements of $SU(2)$ group leads to more simple expressions for higher order terms. The Lagrangian (\ref{eq27}) depend only on $SU(2)$-invariant fields, so the local $SU(2)$ symmetry is factored out of it.
  The concept of generation masses for vector bosons in Electroweak Model  by transformation to radial coordinates    is  further developed in
\cite{MW-10} and \cite{ILM-10}.

The Electroweak Model based on the
nonlinearly realized $SU(2)\times U(1)$ gauge groups was suggested in
\cite{BFQ-08},\cite{BFQ-08-II}. The $SU(2)$ matrix $\Omega$ and the matter field $\Phi$ are taken in the form
\begin{equation}
\Omega = \frac{1}{v}\left(\phi_0+i\sum_{k=1}^3\tau_k\phi_k\right), \quad
\Phi=\left(\begin{array}{c}
 i\phi_1+\phi_2\\
\phi_0-i\phi_3
\end{array} \right), 
\label{dop-7}
\end{equation} 
where $\phi_0=r, \phi_k=r\psi_k, k=1,2,3, v=R$ in our notations (\ref{d1})--(\ref{dop-4}). The nonlinearity of the representation comes from the constraint
\begin{equation}
\Omega^\dagger\Omega=1 \Rightarrow \phi_0^2+\bar{\phi}^2=v^2,
\label{dop-8}
\end{equation} 
which is the same as (\ref{dop-5}).
The quantization of this  model was consistently defined in the perturbative loop-wise expansion and satisfies Physical Unitarity.

The    nonlinear partial-trace $\sigma$-model on $G/H$, which provides mass terms to the intermediate vector bosons associated with the quotient $G/H$ and remain those of $H$ massless, was developed in \cite{ACL-09}. 
An infinite-dimensional symmetry, with non-trivial Noether invariants, which ensures quantum integrability of the model in a non-canonical quantization scheme was found. For $G=SU(2)\times U(1)$ and $H=U_{em}(1)$ this model gives Higgsless alternative to the Standard Model with a partial trace on a quotient manifold $G/H=SU(2)\cong S_3$.

\section{Conclusion}

The  topological mechanism for generation of vector boson masses in the Electroweak Model is discussed. 
Vector boson masses are automatically generated by transformation of the free Lagrangian from the noncompact $R_4$ matter fields space to the compact sphere $ S_3$. This model describes all experimentally observed fields and does not include the (up to now unobserved) scalar Higgs field. $W$- and $Z$-boson masses are expressed through the parameter    $R$ by the same formulas as in the standard case. The free parameter $R$ of the model is now interpreted as the  curvature radius of the spherical matter fields space.
The  development of this topological idea in different aspects, such as transformation to radial coordinates or nonlinearly realized gauge group or nonlinear  partial-trace sigma-model,  is briefly reviewed.

This work is supported by
the program "`Fundamental problems of nonlinear dynamics"' of Russian Academy of Sciences.


\begin{thebibliography}{99}

%
\bibitem{Sh-09}
 Shirkov D V 2009 {\it Usp. Fiz. Nauk} {\bf 179} No. 6 581.
%
%
\bibitem{C-05}
 Casalbuoni R 2005  Deconstructed Higgsless models 
{\it Preprint} hep-ph/0509165 
%
%
\bibitem{CSHKT-04}
 Chivukala R S,  Simmons E H,   He H-J, Kurachi M and  Tanabashi M 2004 
The structure of corrections to electroweak interactions in Higgsless models
{\it Preprint} hep-ph/0406073 
%
\bibitem{S-06-1}
 Slavnov A A 2006 Renormalizable electroweak model without fundamental scalar meson
{\it Preprint} hep-th/0601125 
%
%
\bibitem{CCMT-06}
 Cacciapaglia G,  Cs\'aki C,  Marandella G and  Terning J 2006
A new custodian for a realistic Higgsless model
{\it Preprint} hep-ph/0607156 
%
%
\bibitem{MT-08}
 Moffat J W and  Toth V T 
A finite electroweak model without a Higgs particle
{\it Preprint} arXiv:0812.1991 [hep-th]
%
\bibitem{E-66}
 Efimov G V 1966 {\it Sov. J. Nucl. Phys.} {\bf 2} 138 
%
\bibitem{E-67}
Efimov G V 1967 {\it Sov. J. Nucl. Phys.} {\bf 4} 309 
%
%
\bibitem{Gr-07}
 Gromov N A 2008 
Higgsless Electroweak Model due to the Spherical Geometry 
 {\it J. Phys.: Conf. Series} {\bf 128} 012005  (arXiv:0705.4575 [hep-th])
%
\bibitem{IW-53}
  Inonu E and  Wigner E P 1953
 {\it Proc. Nat. Acad. Sci. USA}  {\bf 39}  510 
 %
\bibitem{NW-93}
 Nappi C R and  Witten E
A WZW model based on non-semisimple group,
{\it Preprint} hep-th/9310112
%
\bibitem{T-95}
 Tsetytlin A A
On gauge theories for non-semisimple groups,
{\it Preprint} hep-th/9505129
%
%
\bibitem{R-99}
Rubakov V A 1999 {\it Classical Gauge Fields}  (Moscow, Editorial URSS) (in Russian)
 %
%
\bibitem{PR-94}
 Pawlowski M and  Raczka R 1994 
 A unified conformal model for fundamental interactions without dynamical Higgs field
 {\it Found. of Phys.} {\bf 24}  1305 ({\it Preprint} hep-th/9407137)
 %
  %
%
\bibitem{F-08}
 Faddeev L D
An alternative interpretation of the Weinberg-Salam model
{\it Preprint} arXiv:0811.3311 [hep-th]
%
\bibitem{MW-10}
 Masson T and  Wallet J - C
  A remark on the spontaneous symmetry breaking mechanism in the Standard Model 
{\it Preprint} arXiv:1001.1176 [hep-th]
%
\bibitem{ILM-10}
 Ilderton A,  Lavelle M and   McMullan D J 2010
 Symmetry breaking, conformal geometry and gauge invariance
 {\it J. Phys. A: Math. Theor.} {\bf 43}, 312002  ({\it Preprint} arXiv:1002.1170 [hep-th]) 
%
%
\bibitem{BFQ-08}
  Bettinelli D,  Ferrari R and  Quadri A 2009
The $SU(2)\otimes U(1)$ Electroweak model based on the nonlinearly realized gauge group   
  {\it Int. J. Mod. Phys. A} {\bf 24}, 2639  
 ({\it Preprint} arXiv:0807.3882 [hep-th]) 
 %
\bibitem{BFQ-08-II}
Bettinelli D,  Ferrari R and  Quadri A 2010 
The $SU(2)\otimes U(1)$ Electroweak model based on the nonlinearly realized gauge group. II
{\it  Acta Phys. Pol. B} {\bf 41} 597  
 ({\it Preprint} arXiv:0809.1994 [hep-th])
%
%
\bibitem{ACL-09}
 Aldaya V,  Calixto M and  Lopez-Ruiz F F 
  A quantizable model of massive gauge vector bosons without Higgs
{\it Preprint} arXiv:0905.3688v1 [hep-th]
%
%
%
%

%
%




\end{thebibliography}
\end{document}